\documentstyle[aps,prl,epsf]{revtex}

\def\be{\begin{equation}}
\def\ee{\end{equation}}
\def\ba{\begin{eqnarray}}
\def\ea{\end{eqnarray}}

\def\C60{A$_x$C$_{60}$}

\def\seX{(TMTSF)$_2$X}

\def\sepf{(TMTSF)$_2$PF$_6$}

\def\secl{(TMTSF)$_2$ClO$_4$}

\def\iTone{T$_1^{-1}$}
\def\iToneT{(T$_1$T)$^{-1}$}

\def\se{$^{77}$Se}
\def\a{{\bf{a}}}
\def\b{{\bf{b'}}}

\begin{document}

\twocolumn[\hsize\textwidth\columnwidth\hsize\csname@twocolumnfalse\endcsname 

\title
{Evidence from \se\ Knight shifts for triplet superconductivity in \sepf}

\author{I.~J.~Lee$^1$, D.~S.~Chow$^2$, W.~G.~Clark$^2$, J. Strouse$^4$, M.~J.~Naughton$^3$, P.~M.~Chaikin$^1$, S.~E.~Brown$^2$}
\address{
1) Department of Physics,
Princeton University
Princeton, N.J. 08544 USA}
\address{
2) Department of Physics and Astronomy, 
UCLA,
Los Angeles, CA 90095-1547 USA}
\address{
3) Department of Chemistry and Biochemistry
UCLA,
Los Angeles, CA 90095-1569 USA}
\address{
4) Department of Physics
Boston College, Chestnut Hill, MA 02167 USA}


\maketitle 

\

\

\

]

{\bf{The layered quasi-one-dimensional molecular superconductor \sepf\ is a very exotic material with a superconducting order parameter whose ground state symmetry has remained ill-defined \cite{Jerome1980,Abrikosov1983,Chaikin1983,Jerome1997}.  Here we present a pulsed NMR Knight shift (K) study of \se\ measured simultaneously with transport in pressurized \sepf. The Knight shift is linearly dependent on the electron spin susceptibility $\chi_s$, and is therefore a direct measure of the spin polarization in the superconducting state. For a singlet superconductor, the spin contribution to the Knight shift, K$_s$, falls rapidly on cooling through the transition. The present experiments indicate no observable change in K between the metallic and superconducting states, and thus strongly support the hypothesis of triplet p-wave superconductivity in \sepf.}} 

The suppression of superconductivity by defects produced by irradiation \cite{Choi1982,Bouffard1982} and by chemical substitution \cite{Coulon1982,Tomic1983} led to early suggestions of p-wave symmetry \cite{Abrikosov1983} as did the presence of a neighboring spin-density-wave phase \cite{Solyom1979}. However, specific heat \cite{Garoche1982} thermal conductivity \cite{Belin1997} and resistive upper critical field studies \cite{Chaikin1983,Murata1987} had indicated conventional, BCS-like pairing. The issue was revived by recent measurements of the upper critical field H$_{c2}$ with substantially improved accuracy in angular alignment and lower temperatures \cite{Lee1995,Lee1997}. Superconductivity persists to field strengths exceeding the Clogston limit \cite{Clogston1962} for singlet superconductors, H$_p$, by several times when the field is applied in the plane of the molecular layers. Even with a field-induced dimensional crossover, which greatly increases the orbital critical field \cite{Lebed1986,Dupuis1993}, an additional mechanism, such as the formation of the inhomogeneous LOFF state \cite{Larkin1965,Fulde1964,Buzdin1987} or triplet superconductivity, is required to exceed H$_p$ in the Bechgaard salts \seX\ (X=PF$_6$, ClO$_4$, AsF$_6$, etc.). The lack of evidence for a first order phase transition between a LOFF state and a uniform superconductor, the observed H$_{c2}>$4H$_p$ and the recent theoretical analysis of the in-plane H$_{c2}$ anisotropy argue against the LOFF state \cite{Lee1997,Lebed1999b}. The present NMR Knight shift study is then strong evidence supporting a triplet state. 

In general, a singlet superconducting ground state leads to a vanishing of the spin contribution, K$_s$, to the total K as T$\to$0. The expected shifts for \se\ are in the range 340-480 ppm for cooling from the normal phase to a singlet superconducting phase. However, we conclude from our measured spectra that no change is observed ($\delta$K$_s$=0$\pm$20 ppm). To ensure that the pressurized sample was superconducting while acquiring the NMR data, we conducted transport measurements time-synchronous with the application of the radiofrequency pulses. 

\begin{figure}[htb]
\epsfxsize=0.9\hsize
\epsffile{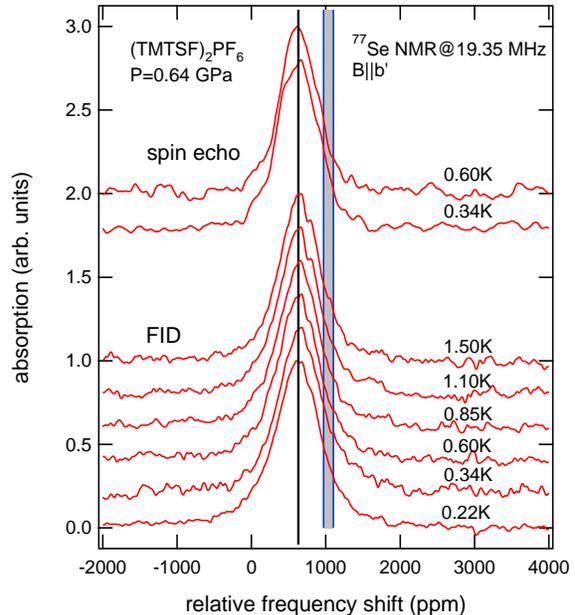}
\caption{\se\ spectra collected at temperatures below and above T$_c$ for the magnetic field B=2.38T oriented parallel to the molecular layers to within 0.1$^\circ$. The solid line marks the measured first moment, and the hashed region marks the expected first moment for a singlet ground state.}
\label{se77prfg1}
\end{figure}

Our principal result is shown in Fig. \ref{se77prfg1}. The lower set of spectra was collected as free induction decays (FIDs), and the upper as spin echoes. Spectra were recorded at temperatures above T$_c$ and below, for a magnetic field aligned parallel to the layers to within 0.1$^{\circ}$ and parallel to the \b-axis to within $\approx$5$^{\circ}$. To within the experimental uncertainty, there is no change in the first moment (marked by the vertical solid line). The vertical hashed region is the estimated range where the center of the spectrum would be if the spin susceptibility had vanished. The lack of any observable difference between the spectra as the temperature is varied indicates the system is not a singlet superconductor. Below, we establish the relationship between the spin susceptibility and the Knight shift.

There is an extensive literature related to NMR work on \sepf, in both the ambient pressure spin-density wave phase \cite{Takahashi1989} and the pressurized metallic phase \cite{Azevedo1984,Creuzet1987}. For the most part, the previous work includes studies of the local magnetic environments of either the protons in the methyl groups or $^{13}$C spin-labeled on the various inequivalent sites. Another possibility is \se. Its abundance is 8\%, I = 1/2, and $^{77}\gamma$=8.129 MHz/T. There are reports of  \se\ spin-lattice relaxation rate measurements in the metallic phase for powdered samples \cite{Azevedo1984} and single crystals \cite{Creuzet1987}, but very little spectroscopy that we know of \cite{Takigawa1986b}. However, it is generally known from band structure calculations as well as EPR studies that the largest spin densities associated with the conduction band are closely linked with molecular orbitals associated the selenium ions.

\begin{figure}[htb]
\epsfxsize=0.9\hsize
\epsffile{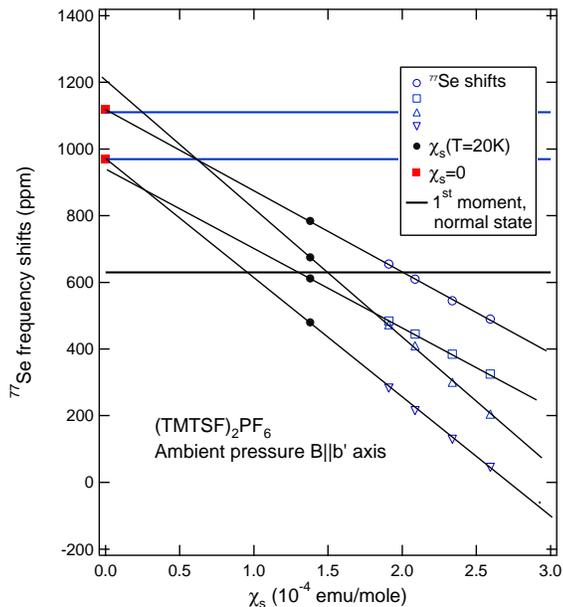}
\caption{NMR shifts K vs. the spin susceptibility $\chi_s$ for magnetic fields applied along the \b-axis. The solid black line is the first moment in the normal state, and the blue lines denote the window for the expected shift at T=0 for a singlet superconductor (the hashed region in Fig. \ref{se77prfg1}.}
\label{se77prfg2}
\end{figure}

Figure \ref{se77prfg2} shows the paramagnetic shifts in the normal state as a function of the spin susceptibility $\chi_s$ for B$\parallel$\b. This type of plot is often used to separate shifts of orbital and chemical origin from shifts originating from the hyperfine coupling, and therefore proportional to $\chi_s$. The absolute value of $\chi_s$ was extracted from Miljak, {\it{et al}} \cite{Miljak1983}, who obtained a strong temperature dependence that probably results from a combination of lattice contraction and low-dimensional correlation effects. Thus, just as for the recent study identifying Sr$_2$RuO$_4$ as a triplet superconductor \cite{Ishida1998}, temperature is a natural implicit parameter relating K to $\chi_s$. The shifts K are measured relative to the NMR line (at $\nu_0$) of  \se\ in Se(CH$_2$)$_2$. That is, K=($\nu-\nu_0)/\nu_0$. We are interested in the expected change $\delta$K upon cooling into the superconducting state from the normal state. The shifts in the normal state at 20K are marked on Fig. \ref{se77prfg2} as K$_s$(T=20K). The extrapolation to $\chi_s$=0 gives the expected shift K($\chi_s$=0) if the superconducting state were singlet. The difference between these two values, K(T=20K) and K($\chi_s$=0), is from the hyperfine coupling to the spin, K$_s$, and it is the expected change $\delta$K for a singlet superconducting ground state. The vertical lines bounding the hashed region in Fig. \ref{se77prfg1} mark the corresponding first moment at 340-480 ppm above the measured value. At the measuring field of B=2.38T, this corresponds to about 6-9 kHz, and we estimate our uncertainty at about 1 kHz. We have also found that \se\ spectroscopy with the field aligned along the \a-axis is much more sensitive. We expect that by changing to that configuration, we can work at much lower fields for comparable or smaller experimental uncertainties.

\begin{figure}[htb]
\epsfxsize=0.9\hsize
\epsffile{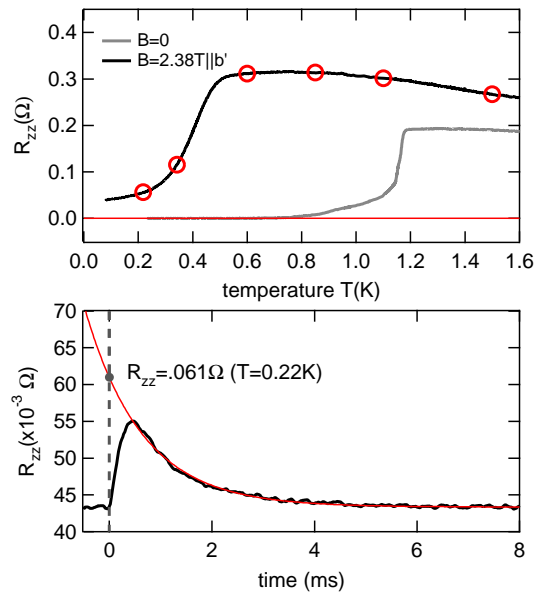}
\caption{a) Interlayer resistance R$_{zz}$ vs. temperature at zero applied magnetic field and at the \se\ measuring field of B=2.38T. b) Time synchronous resistance measurements triggered and recorded simultaneous with the rf pulses for the NMR measurements. The base temperature is T=100mK.} 
\label{se77prfg3}
\end{figure}

In Fig. \ref{se77prfg3}a, we show the interlayer resistance R$_{zz}$ vs. temperature for B=0T, and the \se\ measuring field of B=2.38 T along the \b-axis. There is a sharp reduction at T$_c$(B=0T)=1.18K, followed by a resistive tail (probably related to sample or pressure inhomogeneity) before R$_{zz}$ tends to zero at T = 0.8 K. 

In addition to confirming the superconducting state, {\it{in situ}} transport measurements along with NMR provide a crucial diagnostic tool at the lowest temperatures, since the sample itself can be used as an excellent thermometer in the middle of the superconducting transition. We found that the thermal time constant of the sample (plus surrounding fluids and NMR coils) was approximately 1ms. The NMR spectra are acquired on a time scale of 100$\mu$s.  The duty cycle for the rf pulsing was kept very low, so that the average heating was negligible. A resistance measurement with short time constant made the heating by the rf pulses observable and therefore controllable. 

An example of the time-synchronous transport measurements, recorded under the same conditions as the data which give the spectra in Figure 1, is shown in Fig. \ref{se77prfg3}b. The resistance is measured using a standard four probe lock-in technique at a frequency of 3.14kHz. We verified that the electronic time constant was about 0.3ms. The pulses used for the FID experiments were of duration 1$\mu$s using power levels less than 10mW. (At a recycle time of 1s, the time-averaged power is somewhat less 10nW.) Immediately after the pulse, the sample resistance (temperature) begins to rise, reaching a maximum at about 600$\mu$s later. Afterward, an exponential decay is observed with the time constant of $\approx$1ms. The shape of the heating curve suggests that the NMR coil, not the sample, is the source of the heating. Since the Knight shift measurement is completed in less than 100$\mu$s, it is possible that the sample experiences no heating in this time. The high limiting value of the sample heating is obtained by extrapolating the exponential part of the heating curve back to the time of the pulse. The resistance and temperature at this time are an upper bound to the heating effects. These upper bounds are the temperatures that we quote in the rest of the paper. They are indicated as the open circles in Fig. \ref{se77prfg3}a and are those labeling the spectra in Fig. \ref{se77prfg1}. 

\begin{figure}
\epsfxsize=0.9\hsize
\epsffile{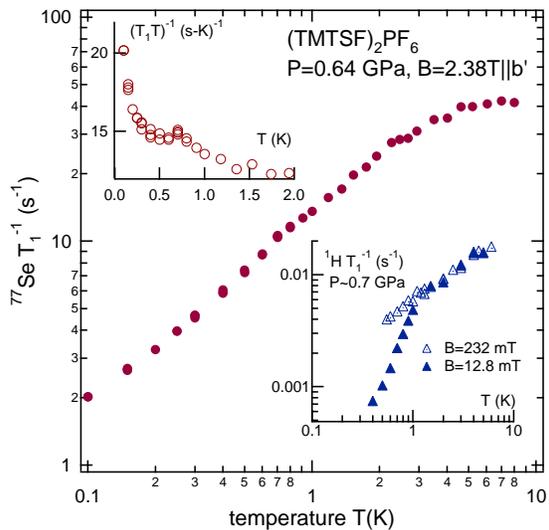}
\caption{$^{77}$Se T$_1^{-1}$ vs. temperature for B=2.38T $\parallel$ \b. In the lower inset are shown the results of field-cycled $^1$H \iTone\ measurements; at the top, the detailed temperature dependence of \iToneT\ near the superconducting transition is depicted.}
\label{se77prfg4}
\end{figure}

To obtain an independent and bulk measure of the superconducting transition, we recorded the spin-lattice relaxation rates \iTone\ for \se\ that are shown in Fig. \ref{se77prfg4}. Plotted in this way, the results look generally similar to what is expected for the normal state of pressurized \sepf\ \cite{Creuzet1987,Azevedo1984}. However, a distinct peak that we associate with the onset of superconductivity is evident in \iToneT\ near to T=0.7K (see upper inset). Ordinarily, we expect that the signature from transport would occur at the same temperature as for \iTone, although the transport results of Fig. \ref{se77prfg3}a indicate a sharp resistance drop at T$\sim$0.5K. We note that dynamic processes related to flux motion (when the sample is cooled in a magnetic field) can influence both resistance and T$_1$ measurements. 

A related issue is the general tendency for the relaxation rate to approach the normal state value. Well below T$_c$(H) in a strong Type II superconductor, \iTone$\sim$~TH/H$_{c2}$ because of the normal fraction in the vortex cores and the Korringa relation \cite{MacLaughlin1976}. In fact, \iTone\ for \sepf\ appears to be extraordinarily sensitive to magnetic field, which we demonstrate by way of field-cycling experiments using the methyl group protons as a probe. Shown in the lower inset to Fig. \ref{se77prfg4} are experiments on another sample with slightly higher pressure at B=12.8mT and B=232mT. The behavior at the lower field is similar to that observed by Takigawa, {\it{et al}} \cite{Takigawa1987}, for field-cycled $^1$H relaxation in superconducting \secl: there is no indication for a Hebel-Slichter peak, and below the transition \iTone$\sim$T$^{\beta}$, with $\beta$=2.5 (in Ref. \cite{Takigawa1987}, $\beta$ was reported to be 3.0). At the resonance field of 232mT, the fast drop of \iTone\ is nearly absent, although there is a distinct change in slope at T=1K. Remarkably, the applied field B=232mT at T=0.5K is far below characteristic values for H$_{c2}$ observed by Lee, {\it{et al}} \cite{Lee1997} or in earlier studies. The important points are these: 1) We have a clear signature for superconductivity from the \se\ relaxation rates, and 2) even though the rates are high in the superconducting state they are not simply related to the volume fraction of the normal cores. 

The evidence for no change in spin susceptibility between the normal and superconducting states would be consistent with the so-called Anderson-Brinkman-Morel state identified with superfluid $^3$He-A phase \cite{Leggett1983}, where there are spin-up and spin-down pairs but no pairs with S$_z$ = 0. Lebed has proposed an order parameter with the spins oriented parallel to the \b-axis to account for the findings here and the critical field anisotropy. In that case, a change in shift is expected for fields applied parallel to the \a-axis. Experiments designed to verify that prediction are underway, along with a search for the longitudinal resonance \cite{Leggett1983} which gives a measure of the order parameter of the condensed triplet phase. 


ACKNOWLEDGEMENTS. The authors thank Hae-Young Kee and Andrei Lebed for many discussions on this topic. The work was supported in part by the National Science Foundation.

\end{document}